\documentclass[twoside,12pt]{article}  
\usepackage{CJK}   
\usepackage{indentfirst}  
\usepackage{bm}    
\usepackage{amsmath}

\usepackage{graphicx}
\usepackage{subfigure}
\usepackage{amsmath,amsfonts,amssymb,graphicx}
\usepackage{appendix}

\usepackage{graphicx}  

\begin{document}    

\begin{center}
\LARGE\bf Anomalous  quantum mechanics $^{*}$   
\end{center}

\footnotetext{\hspace*{-.45cm}\footnotesize $^*$Project supported by
the National Natural Science Foundation of China (Grant No.
11626047) and the Foundation for Young Key Teachers of Chengdu
University of Technology, China (Grant No. KYGG201414).}
\footnotetext{\hspace*{-.45cm}\footnotesize $^\dag$Corresponding
author. E-mail: zhanghong123@cdut.cn (Zhang H)
 }

\begin{center}
\rm Hong Zhang$^{\dagger}$, \ \
\end{center}

\begin{center}
\begin{footnotesize} \sl
Department of Mathematics Teaching, Chengdu University of
Technology, Chengdu 610059, China
\end{footnotesize}
\end{center}

\begin{center}
\footnotesize (Received xx xx; revised manuscript received xx
xx )
\end{center}

\vspace*{2mm}

\begin{center}
\begin{minipage}{15.5cm}
\parindent 20pt\footnotesize
The fractional operators together with exponential quantum in
coordinate and momentum space corresponding to the power of
observables are introduced. Based on an exponential relation between
energy and momentum, the fractional Schr$\ddot{\texttt{o}}$dinger
equations for the free particle and the one in potential fields in
heterogeneous complex media are found. The fractional equation of
motion
 and the fractional virial theorem for anomalous  quantum mechanics are then developed. Applying the fractional virial
 theorem, we derive an anomalous hydrogen atom whose transition energy values are much higher  than that of Bohr hydrogen atom. The anomalous Heisenberg picture being
equivalent to the fractional Schr$\ddot{\texttt{o}}$dinger picture
is also discussed.
\end{minipage}

\end{center}

\begin{center}
\begin{minipage}{15.5cm}
\begin{minipage}[t]{2.3cm}{\bf Keywords:}\end{minipage}
\begin{minipage}[t]{13.1cm}
Anomalous operator,Fractional virial theorem, Anomalous Heisenberg
picture

\end{minipage}\par\vglue8pt
{\bf PACS: }
03.65.-w, 05.30.-d, 31.10.+z
\end{minipage}
\end{center}

\section{Introduction}
 In classical mechanics, the general requirements
imposed by the homogeneity and isotropy of space and by Galileo's
relativity principle lead to a quadratic dependence of the energy of
the particle on its momentum: $E = p^2/2m$, where the constant $m$
is the mass of the particle.$^{[1-4]}$ However, most of the actual
systems are nonhomogeneous and in which the relations $E = p^2/2m$
does not still hold for the energy and momentum. In recent years
anomalous properties in disordered systems and the nonquadratic
dependences of the energy on momentum attracted growing
attention.$^{[5-15]}$

In 2000, Laskin first discussed the anomalous properties in quantum
physics and derived the fractional Schr$\ddot{\texttt{o}}$dinger
equation with anomalous exponent $1<\alpha\leq 2$ by applying the
path integrals over the L$\acute{e}$vy paths.$^{[10]}$

The approach to quantum dynamics describing the state of a particle
by state function $\varphi(x,t)$ is the
Schr$\ddot{\texttt{o}}$dinger equation, known as the
Schr$\ddot{\texttt{o}}$dinger picture. Another formulation
of quantum dynamics is the Heisenberg picture where observables or
the corresponding operators depend on time and
obey the Heisenberg equation of motion.$^{[16]}$
  The
Heisenberg picture is a particularly useful device in relativistic
quantum mechanics.$^{[17]}$

In this article we shall introduce anomalous quantum mechanics by
introducing fractional operators and exponential quanta. Besides,
based on exponential hypothesis that in heterogeneous complex media
the energy $E$ and momentum $p$ of a particle of mass $m$ satisfy
$E=\frac{p^{2\alpha}}{(2m)^{\alpha}}$,
 we derived the fractional Schr$\ddot{\texttt{o}}$dinger equation for the free particle
and the
 one in the potential $V(x^{\beta})$.  We also derive the fractional equation of
motion
 and the fractional virial theorem for anomalous  quantum mechanics.
 Moreover, we give an anomalous hydrogen atom whose transition energy values are much higher and can match up the anomalous
$X-$ray spectrums observed by Wang, etc.. Finally, we will discuss
anomalous Heisenberg picture being equivalent to the fractional
Schr$\ddot{\texttt{o}}$dinger equation.

\section{Fractional operators and exponential  quanta}

\subsection{Anomalous operators in coordinate space}

 We start by introducing the fractional operators and
exponential   quanta in coordinate  space for anomalous quantum
mechanics. For simplicity, in what follows we shall consider a
particle moving in $x$ direction.

For a state normalized according to condition $\int
\psi^{*}(x,t)\psi(x,t)dx=1$, the average value of the power of
coordinate $x^{\alpha}$ with the exponent $\alpha>0$ is
 \begin{equation}\langle x^{\alpha}\rangle=\int
\psi^{*}(x,t)x^{\alpha}\psi(x,t)dx
\end{equation}
where $\psi^{*}(x,t)\psi(x,t)=\mid\psi(x,t)\mid^{2}$ denotes the
probability of the various values of coordinate, and $x^{\alpha}$ is
regarded as the operator corresponding to $x^{\alpha}$  that
multiplies the wave function by $x^{\alpha}$,
 that is,
\begin{equation}
\hat{x^{\alpha}}=x^{\alpha}.
\end{equation}

We now consider the operator $\hat{p^{\alpha}}$ corresponding to the
power of momentum $p^{\alpha}$ in coordinate space.  Note that this
operator must be defined so that the average value of $p^{\alpha}$
can be represented in the
form 
\begin{equation}
\langle p^{\alpha}\rangle=\int\psi^{*}(x,t)
\hat{p^{\alpha}}\psi(x,t)dx.
\end{equation}
On the other hand, this average
value is determined from the momentum wave function by
\begin{equation}\langle
p^{\alpha}\rangle=\int\phi^{*}(p,t)p^{\alpha}\phi(p,t) dp.
\end{equation}
where $\phi(p,t)$ is
related to $\psi(x,t)$ by the reciprocal Fourier transforms:
\begin{equation}
\psi(x,t)=\frac{1}{(2\pi h)^{1/2}}\int
\phi(p,t)\exp(\frac{i}{h}p\cdot x)dp,
\end{equation}
and
\begin{equation}
\phi(p,t)=\frac{1}{(2\pi h)^{1/2}}\int
\psi(x,t)\exp(-\frac{i}{h}p\cdot x)dx.
\end{equation}
Substituting the expression (6) for  $\phi^{*}(p,t)$, we have
\begin{equation}
\langle p^{\alpha}\rangle= \int \int\psi^{*}(x,t)\frac{1}{(2\pi
h)^{1/2}} \exp(\frac{i}{h}p\cdot x)p^{\alpha}\phi(p,t)dxdp.
\end{equation}
Let us write
$$\exp(\frac{i}{h}p\cdot
x)p^{\alpha}=[(-ih)^{\alpha}D_x^{\alpha}]\exp(\frac{i}{h}p\cdot
x),$$ where $D_x^{\alpha}$ is a type of fractional
derivative$^{[18]}$, satisfying
\begin{equation}
D_x^{\alpha}\exp(ax)=a^{\alpha}\exp(ax),
\end{equation}
and  integration by parts formula
\begin{equation}
(\varphi, D_x^{\alpha}\psi)=(D_x^{\alpha}\varphi, \psi).
\end{equation}
The integral (7) then becomes \begin{equation}\langle
p^{\alpha}\rangle=\int \int\psi^{*}(x,t)\frac{1}{(2\pi
h)^{1/2}}[(-ih)^{\alpha}D_x^{\alpha}]\exp(\frac{i}{h}p\cdot
x)\phi(p,t)dxdp.\end{equation} Using Eq.(5), we find
\begin{equation}
\langle p^{\alpha}\rangle=\int\psi^{*}(x,t)
[(-ih)^{\alpha}D_x^{\alpha}]\psi(x,t)dx.
\end{equation}
Comparing this with (3), we see that the operator of the power of
momentum $p^{\alpha}$ in the coordinate representation is
\begin{equation}
\hat{p^{\alpha}}=(-ih)^{\alpha}D_x^{\alpha},
\end{equation}
which is a  a fractional operator times exponential quantum. When
the power $\alpha$ is a positive integer, the operator
$(-ih)^{\alpha}D_x^{\alpha}$ reduces to the ordinary operator
$\underline{}(-ih\frac{\partial}{\partial x})^{n}$. More generally,
the same technique gives for analytic function of $p^{\alpha}$ that
can be expressed as a power series $f(p^{\alpha})=\Sigma C_n
(p^{\alpha})^n$ where $n$ denotes positive integer, then yields
\begin{equation}\langle f(p^{\alpha})\rangle=\int\psi^{*}(x,t)
\{\Sigma
C_n[(-ih)^{\alpha}D_x^{\alpha}]^n\}\psi(x,t)dx,\end{equation} from
which we have the function $f(p^{\alpha})$ has a corresponding
operator $f(\hat{p^{\alpha}})$, which we call anomalous operator in
coordinate space.

\subsection{Anomalous operators in momentum space }
We now propose the fractional operators and exponential quanta in
momentum space for anomalous quantum mechanics. One can find that
the operator corresponding to $p^{\alpha}$ in momentum space is
represented as $p^{\alpha}$ for $\alpha>0$, since
 the expectation value of the power of momentum $p^{\alpha}$ is
$$\langle p^{\alpha}\rangle=\int \phi^{*}(p,t)p^{\alpha}\phi(p,t) dp.$$

In analogy with the evaluation of $p^{\alpha}$ in $x$ space, we get
$$\langle x^{\alpha}\rangle=\int\psi^{*}(p,t)
[(ih)^{\alpha}D_p^{\alpha}]\psi(p,t)dp,$$ which implies that the
operator corresponding to $x^{\alpha}$ in momentum space is a
fractional operator times exponential quantum
\begin {equation}
x^{\alpha}=(i h)^{\alpha}D_p^{\alpha}
\end {equation}
 Here $D_p^{\alpha}$ is a type of fractional derivative,
satisfying
\begin{equation}
D_p^{\alpha}\exp(ap)=a^{\alpha}\exp(ap),
\end{equation}
and  integration by parts formula
\begin{equation}
(\varphi, D_p^{\alpha}\psi)=(D_p^{\alpha}\varphi, \psi).
\end{equation}
If $f(x^{\alpha})=\sum C_n (x^{\alpha})^n$, it can then be shown
that
$$\langle f(x^{\alpha})\rangle=\int\phi^{*}(p,t)
\{\sum C_n[(ih)^{\alpha}D_p^{\alpha}]^n\}\phi(p,t)dp,$$ from which
we get
$$\hat{f[x^{\alpha}]}=f[(ih)^{\alpha}D_p^{\alpha}].$$
We call this generalized operator, anomalous operator in momentum
space.

\subsection{Linearity and Hermiticity of anomalous operators }

Because of
$$(-ih)^{\alpha}D_x^{\alpha}(\psi_1(x)+\psi_2(x))=
(-ih)^{\alpha}D_x^{\alpha}(\psi_1(x))+(-ih)^{\alpha}D_x^{\alpha}(\psi_2(x)),$$
and
$$(-ih)^{\alpha}D_x^{\alpha}(C\psi(x))=C(-ih)^{\alpha}D_x^{\alpha}(\psi(x)),$$
 we obtain that the operator $\hat{p^{\alpha}}=(-ih)^{\alpha}D_x^{\alpha}$ is a linear operator. Here $C$ is an arbitrary constant.

We now consider the transposed operator
$$\tilde{(-ih)^{\alpha}D_x^{\alpha}}=(ih)^{\alpha}D_x^{\alpha}.$$
According to the definition of transposed operator,
we get
$$\int_{-\infty}^{+\infty}\varphi^{*}(x,t)(-ih)^{\alpha}D_x^{\alpha}\psi(x,t)dx
=\int_{-\infty}^{+\infty}\psi(x,t)(ih)^{\alpha}D_x^{\alpha}\varphi^{*}(x,t)dx$$
\begin{equation}
=\int_{-\infty}^{+\infty}((-ih)^{\alpha}D_x^{\alpha})\varphi(x,t))^{*}\psi(x,t)dx,
\end{equation}
where the superscript $*$ denotes the complex conjugate. Thus,
 $$(\tilde{(-ih)^{\alpha}D_x^{\alpha}})^{*}=(-ih)^{\alpha}D_x^{\alpha}.$$
This means that the operator
$\hat{p^{\alpha}}=(-ih)^{\alpha}D_x^{\alpha}$ is a Hermitian or
self-adjoint operator. We also have any anomalous operator
$\hat{O^{\alpha}}$ that can be expressed as a power series
$\sum_{i=1}^{n}C_n(\hat{p^{\alpha}})^{n}$ is Hermitian.

Similarly, one can obtain that the operator $x^{\alpha}$ and the
function $\hat{f(x^{\alpha})}$ in momentum space is linear and
Hermitian.

\section{Fractional Schr$\ddot{\texttt{o}}$dinger equation}

 We
assume that in heterogeneous complex media the energy $E$ and
momentum $p$ of a particle of mass $m$ are related by an exponential
relation
\begin{equation}
E=\frac{p^{2\alpha}}{(2m)^{\alpha}} .
\end{equation}
where the constant $m$ is the mass of the particle.
 Combining Eq.(18) with
the de Broglie hypothesis
$$ P=hv/c,$$ and
the Einstein relation
$$E=hv,$$ where $v$ denotes the frequency of the incident,
$c$ is the velocity of light, $h$ is Planck's constant, we obtain
the complete (time-dependent) wave function as
\begin{equation}
\psi(x,t)=Ce^{i/h(px-\frac{p^{2\alpha}}{(2m)^{2\alpha}}t)},
\end{equation}
where $C$ is a constant.
 Each such function, a
plane wave, describes a state in which the free particle has a
definite energy $\frac{p^{\alpha}}{(2m)^{\alpha}}$ and momentum $p$.
We now look for a differential equation having (19) as its solution.
We have from (19)
\begin{equation}
ih\frac{\partial \psi(x,t)}{\partial
t}=\frac{p^{2\alpha}}{(2m)^{\alpha}}\psi(x,t),
\end{equation}
\begin{equation}
(-ih)^{\alpha}D_x^{\alpha}\psi(x,t)=p^{\alpha}\psi(x,t),
\end{equation}
 so what we can write the differential equation
 \begin{equation}
ih\frac{\partial \psi(x,t)}{\partial t}=
\frac{((-ih)^{\alpha}D_x^{\alpha})^2}{(2m)^{\alpha}}\psi(x,t).
\end{equation}
Similarly, we can find Eq.(22) for a free particle with a wave
packet,
$$\psi(x,t)=\frac{1}{(2\pi h)^{1/2}}\int
\phi(p,t)e^{i/h(px-\frac{p^{2\alpha}}{(2m)^{2\alpha}}t)}dp.$$

A generalization of the free particle wave equation to the case of
motion of a particle acted by a force given by a potential energy
function $V(x^{\beta})$, is the fractional Schr$\ddot{o}$dinger
equation:
\begin{equation}
ih\frac{\partial \psi(x,t)}{\partial t}=
\frac{((-ih)^{\alpha}D_x^{\alpha})^2}{(2m)^{\alpha}}\psi(x,t)+V(x^{\beta})
\psi(x,t).
\end{equation}
Noting that
$((-ih)^{\alpha}D_x^{\alpha})^2=\hat{(p^{\alpha})^2}=\hat{p^{2\alpha}}$,
we can write Eq.(23) in the form
\begin{equation}
ih\frac{\partial \psi(x,t)}{\partial t}=\hat{H_{\alpha,\beta}}
\psi(x,t),
\end{equation}
 where $\hat{H_{\alpha,\beta}}$ represents the corresponding
operator of  total energy or Hamiltonian
$$H_{\alpha,\beta}=T+V=\frac{p^{2\alpha}}{(2m)^{\alpha}}+V(x^{\beta}).$$

By the Hermiticity property, in combination with the fact that
$\hat{p^{\alpha}}$ is Hermitian, we have the Hamiltonian
$\hat{H_{\alpha,\beta}}$ is also a Hermitian operator.

If the system is conservative and the Hamiltonian operator
$H_{\alpha,\beta}$ does not depend on time $t$, the solution of
$(24)$ may be written as
\begin{equation}
\psi(x,t)=e^{-\frac{iEt}{h}}\psi_{E}(x).
\end{equation}
A particle in this state of this type has a well defined energy E,
since $E/h$ is the time rate of change of the phase of the wave
function. Substituting (25) into (24) we find the time-independent
equation fractional Schr$\ddot{\texttt{o}}$dinger equation
$$\hat{H_{\alpha,\beta}}\psi_{E}(x)=E\psi_{E}(x).$$
or, more explicitly,
$$\frac{((-ih)^{\alpha}D_x^{\alpha})^2}{(2m)^{\alpha}}\psi(x,t)+V(x^{\beta})
\psi(x,t)\psi_{E}(x)=E\psi_{E}(x).$$ For only certain values of $E$
will this equation have solutions. These values are the possible
energy values of the particle, and the corresponding $\psi_{E}(x)$
is the wave function of the particle when it has energy E.

\section{Fractional equation of motion }

We then derive the fractional equation of motion for anomalous
quantum mechanics.

 From the expectation value of one
anomalous operator
\begin{equation}
\langle\hat{O_{\alpha}}\rangle=(\psi(x,t),
\hat{O_{\alpha}}\psi(x,t)),
\end{equation}
we have
\begin{equation}
ih\frac{d}{dt}\langle\hat{O_{\alpha}}\rangle=ih(\frac{\partial
\psi(x,t)}{\partial t}, \hat{O_{\alpha}}\psi(x,t))+ih(\psi(x,t),
\hat{O_{\alpha}}\frac{\partial \psi(x,t)}{\partial t})+ih(\psi(x,t),
\frac{\partial \hat{O_{\alpha}}}{\partial t}\psi(x,t)).
\end{equation}
From Eq.(24), we find
\begin{equation}
ih\frac{d}{dt}\langle\hat{O_{\alpha}}\rangle=-(\hat{H_{\alpha,\beta}}\psi(x,t),
\hat{H_{\alpha,\beta}}\psi(x,t))+(\psi(x,t),
\hat{O_{\alpha}}\hat{H_{\alpha,\beta}}\psi(x,t))+ih\langle\frac{\partial
\hat{O_{\alpha}}}{\partial t}\rangle.
\end{equation}
Since the operator $\hat{H_{\alpha,\beta}}$ is Hermitian, one
obtains
$$(\hat{H_{\alpha,\beta}}\psi(x,t),
\hat{O_{\alpha}}\psi(x,t))=(\psi(x,t),
\hat{H_{\alpha,\beta}}\hat{O_{\alpha}}\psi(x,t)).$$ Therefore,
$$ih\frac{d}{dt}\langle\hat{O_{\alpha}}\rangle=\langle\hat{O_{\alpha}}\hat{H_{\alpha,\beta}}-\hat{H_{\alpha,\beta}}\hat{O_{\alpha}}\rangle
+ih\langle\frac{\partial \hat{O_{\alpha}}}{\partial t}\rangle$$
\begin{equation}
=\langle[\hat{O_{\alpha}},\hat{H_{\alpha,\beta}}]\rangle
+ih\langle\frac{\partial \hat{O_{\alpha}}}{\partial t}\rangle.
\end{equation}
In the second series we used the fractional poisson bracket with
exponential  quanta
$$[\hat{O_{\alpha}},\hat{H_{\alpha,\beta}}]=\hat{O_{\alpha}}\hat{H_{\alpha,\beta}}-\hat{H_{\alpha,\beta}}\hat{O_{\alpha}}.$$

Noting that
$\langle\hat{O_{\alpha}}\rangle=\int\psi^{*}(x,t)\hat{O_{\alpha}}\psi(x,t)
dx$ is independent of the position $x$, we consider
\begin{equation}
\frac{d}{dt}\langle\hat{O_{\alpha}}\rangle=
\langle\frac{d}{dt}\hat{O_{\alpha}}\rangle=
\langle\int\psi^{*}(x,t)\frac{d\hat{O_{\alpha}}}{dt}\psi(x,t)
dx\rangle.
\end{equation}
Eq.(29) then becomes
\begin{equation}
ih\frac{d}{dt}\hat{O_{\alpha}}=[\hat{O_{\alpha}},\hat{H_{\alpha,\beta}}]
+ih\frac{\partial \hat{O_{\alpha}}}{\partial t},
\end{equation}
which is the fractional equation of motion for  anomalous quantum
mechanics. If the operator $\hat{O_{\alpha}}$ is independent of time
and $\alpha$ is integer, then Eq.(31) reduces to the usual equation
of motion
\begin{equation}
ih\frac{d}{dt}\hat{O_{\alpha}}=
[\hat{O_{\alpha}},\hat{H_{\alpha,\beta}}].
\end{equation}

\section{Fractional virial theorem }
We will further derive the fractional virial theorem for the
particle with Hamiltonian
\begin{equation}
H_{\alpha,\beta}=T+V=\frac{p^{2\alpha}}{(2m)^{\alpha}}+V(x^{\beta}).
\end{equation} By Eq.(32), one gets
$$ih\frac{d}{dt}\langle\hat{x}\cdot \hat{p}\rangle=
\langle[\hat{x}\cdot \hat{p},\hat{H_{\alpha,\beta}}]\rangle.$$
Applying
the rules of commutator bracket notation,
we obtain
$$\frac{d}{dt}\langle\hat{x}\cdot \hat{p}\rangle=\frac{1}{ih}\{
\langle[\hat{x}\cdot
\hat{p},\frac{p^{2\alpha}}{(2m)^{\alpha}}]\rangle+
\langle[\hat{x}\cdot \hat{p},V(x^{\beta})]\rangle\}.$$
$$=\frac{1}{ih}\{ 2ih\alpha\langle
\frac{p^{2\alpha}}{(2m)^{\alpha}}\rangle-ih \langle\hat{x}\cdot
\frac{\partial}{\partial x}V(x^{\beta})\rangle\}.$$
\begin{equation}
=2\alpha\langle \frac{p^{2\alpha}}{(2m)^{\alpha}}\rangle- \langle
x\cdot \frac{\partial}{\partial x}V(x^{\beta})\rangle.
\end{equation}
For a stationary state all expectation values in (34) are constant in
time, and it follows that
\begin{equation}
2\alpha\langle \hat{T}\rangle= \langle x\cdot
\frac{\partial}{\partial x}V(x^{\beta})\rangle,
\end{equation}
 where $\hat{T}$ denotes the anomalous operator for kinetic energy. We call Eq.(34) and (35) the fractional virial theorems.
 When $\alpha=1$, $\beta=1$, Eq.(35) yields the classical virial theorem
\begin{equation}
2\langle \frac{\hat{p^{2}}}{2m}\rangle= \langle x\cdot
\frac{\partial}{\partial x}V(x)\rangle.
\end{equation}

\section{Anomalous hydrogen atom}

We will consider one kind of anomalous hydrogen atom whose total
energy is
$$H_{\alpha,\beta}=\frac{p^{2\alpha}}{(2m)^{\alpha}}-\frac{e^2}{4\pi\varepsilon_0 r},,$$
where $\frac{p^{2\alpha}}{(2m)^{\alpha}}$ is the kinetic energy, and
$-\frac{e^2}{4\pi\varepsilon_0 r}$ is the potential energy.

According to the fractional virial theorem (35), we obtain
$$2\alpha\frac{p^{2\alpha}}{(2m)^{\alpha}}= \frac{e^2}{4\pi\varepsilon_0 r}.$$
Using the  first Niels Bohr's postulate $pa_n=n\hbar$ , one gets
\begin{equation}
\frac{2\alpha}{(2m)^{\alpha}}(\frac{n\hbar}{a_n})^{2\alpha}=\frac{e^2}{4\pi\varepsilon_0
a_n},
\end{equation}
 where $a_n$ denotes circular orbits of radius. From (37), we
find
\begin{equation}
a_n=[\frac{8\pi\varepsilon_0
\hbar^{2\alpha}\alpha}{e^2(2m)^{\alpha}}]^{1/(2\alpha-1)}
n^{2\alpha/(2\alpha-1)}.
\end{equation}
Thus, the total energy levels of the anomalous hydrogen atom is
\begin{equation}
H_{\alpha,\beta}=(1-2\alpha)(\frac{me^4}{8h^2\varepsilon_0^2\alpha
^2})^{\alpha/(2\alpha-1)} n^{-2\alpha/(2\alpha-1)}.
\end{equation}
When $\alpha=1$, we find the well-known results of the Bohr theory:
\begin{equation}a_n=\frac{4\pi\varepsilon_0 \hbar^{2}}{e^2
m}n^{2}=0.529\times 10^{-10}m,\end{equation} and
\begin{equation}E_n=-\frac{me^4}{8h^2\varepsilon_0^2} n^{-2}=13.6ev\end{equation}

Let us now consider a transition from $k$ to $n (k>n)$. According to
Einstein's relation $\bigtriangleup E=hv$, we get the transition
energy
\begin{equation}
\bigtriangleup
E_{kn}=hv=(2\alpha-1)(\frac{me^4}{8h^2\varepsilon_0^2\alpha^2})^{\alpha/(2\alpha-1)}
[n^{-2\alpha/(2\alpha-1)}-k^{-2\alpha/(2\alpha-1)}].
\end{equation}
Let $\beta=2\alpha$. Then Eq.(42) can be written as $$\bigtriangleup
E_{kn}=(\beta-1)(\frac{2\sqrt{13.6\times 1.6\times
10^{-19}}}{\beta})^{\beta/(\beta-1)}[n^{-\beta/(\beta-1)}-k^{-\beta/(\beta-1)}]$$
 Assuming $\beta=2.3566$, we find
$$\bigtriangleup E_{2,1}=2.0357kev$$
$$\bigtriangleup E_{3,1}=2.4767kev$$
$$\bigtriangleup E_{4,1}=2.6464kev$$
$$\bigtriangleup E_{20,1}=2.8920kev$$
$$\bigtriangleup E_{120,1}=2.9073kev$$
which are much higher than what are usual for the Bohr atom. And
this theoretical results are in good agreement with the anomalous
experimental data detected by Wang, etc. for one new hydrogen
atom.$^{[19-21]}$ Note also that the first three radiuses of the
anomalous hydrogen atom are as following:
$$~~a_1=1.8613\times 10^{-16}m,$$
$$a_2= 6.205\times 10^{-16}m,$$
$$~a_3= 1.2550\times 10^{-15}m,$$
which are  much smaller than the Bohr radius.

\section{Anomalous Heisenberg  picture. }

\subsection{Matrix representation of an anomalous operator}

When a basis is given, an anomalous operator $\hat{O_{\alpha}}$ in a
given representation can be characterized by its effect on the basis
vectors $\{\psi_n\}$. Indeed, being a vector in the form of series
$\psi=\sum_{n} a_n\psi_n$ in the space, $\hat{O_{\alpha}}\psi$ can
obviously be expanded as
\begin{equation}
\hat{O_{\alpha}}\psi_j=\sum_{i}\psi_i O_{ij},
\end{equation}
where
\begin{equation}
O_{ij}=(\psi_i, \hat{O_{\alpha}}\psi_j)
\end{equation}
 which, owing to the linearity of $\hat{O_{\alpha}}$, completely specify the effect
of $\hat{O_{\alpha}}$ on any vector $\psi$. To see this explicitly,
we note,
\begin{equation}
\phi=\hat{O_{\alpha}}\psi=\hat{O_{\alpha}}\sum_{j}a_j\psi_j
=\sum_{j}a_j\hat{O_{\alpha}}\psi_j =\sum_{j}\sum_{i}a_j\psi_i O_{ij}
=\sum_{i}\psi_i(\sum_{j} O_{ij} a_j).
\end{equation}
 Since
$\phi=\sum_{i}b_i\psi_i$, we find
\begin{equation}
b_i=\sum_{j} O_{ij} a_j,
\end{equation}
proving the contention that the effect of $\hat{A_{\alpha}}$ on any
vector is known if all $O_{ij}$ are known.

\subsection{Anomalous Heisenberg  picture}

Let us define a time  translation operator
\begin{equation}
U(t,t_0)=e^{-\frac{i}{h}\hat{H_{\alpha,\beta}}(t-t_0)},
\end{equation}
where $\hat{H_{\alpha,\beta}}$ is an anomalous Hamiltonian of the
system. Using the Dirac notation, we then express the state vector
$|\psi(x,t)\rangle$ in terms of $|\psi(x,t_0)\rangle$ by the
relation: $|\psi(x,t)\rangle=U(t,t_0)|\psi(x,t_0)\rangle$. Note that
the time  translation operator $U(t,t_0)$ is unitary, it is
sufficient to perform the unitary transformation associated with the
operator $U^{\dag}(t,t_0)$ to obtain a constant transformed vector
\begin{equation}
U^{\dag}(t,t_0)|\psi(x,t)\rangle=|\psi(x,t_0)\rangle.
\end{equation}
Every operator  $\hat{O_{\alpha}}$ can be transformed into
time-dependent operator
\begin{equation}
\hat{O_{\alpha}^H}=U^{\dag}(t,t_0)\hat{O_{\alpha}}U(t,t_0).
\end{equation}
Differentiating the expression (49) with respect to time, we obtain
\begin{equation}
\frac{\partial}{\partial t}\hat{O_{\alpha}^H}
=\frac{i}{h}[\hat{H_{\alpha,\beta}},
\hat{O_{\alpha}^H}]+\frac{\partial
 \hat{O_{\alpha}^H}}{\partial t},
\end{equation}
where the last term is vanishing  when the operators
$\hat{O_{\alpha}}$ does not  depend on $t$. Note that Eq.(50) is
similar in form to (31) but has a somewhat different significance:
the expression (31) defines the operator corresponding to the
physical quantity $\langle \hat{O_{\alpha}}\rangle$, while the
left-hand side of equation (50) is the time derivative of the
operator of the quantity $ \hat{O_{\alpha}^H}$ itself.

Let $\psi=\Sigma_i a_i(t)\Psi_i$ be the expansion of an arbitrary
wave function in terms of the  basis $\Psi_i$ in coordinate
representation. If we substitute this expansion in Eq.(23) for free
particle, we obtain
\begin{equation}
ih\sum_i \dot{a}_i(t)\Psi_i= \sum_i
a_i(t)\frac{((-ih)^{\alpha}D_x^{\alpha})^2}{(2m)^{\alpha}} \Psi_i.
\end{equation}
Now multiplying this equation by $\psi_j^{\ast}$, and using normalization and orthogonality of the $\Psi_j$, we then find
\begin{equation}
ih \dot{a}_j(t)\Psi_i= \sum_i H_{ji} a_i(t),
\end{equation}
where
$$H_{ji}=(\Psi_j,
\frac{((-ih)^{\alpha}D_x^{\alpha})^2}{(2m)^{\alpha}} \Psi_i).$$ This
equation completely defines how the $a_i(t)$ change, whenever the
$a_i(t)$ are known at anyone time.

\subsection{Fractional secular equation}

 We now consider the the eigenvalue equation in a given
representation as
\begin{equation}
\hat{O_{\alpha}}\phi=\lambda \phi,
\end{equation}
 where $\phi=\sum_i c_i\psi_i$ is an eigenvector in terms of the  basis $\psi_i$ with the  eigenvalue $\lambda$.
 Since the  operator
$\hat{O_{\alpha}}$ is  Hermitian, any eigenvalue $\lambda$ of
$\hat{O_{\alpha}}$ is a real number. Substituting the matrix
representation of $\hat{O_{\alpha}}$ and $\phi$
 in the equation (53), we get
 $$\sum_{j}O_{ij}c_{j}=\lambda c_{i}$$
 or
 $$\sum_{j}(O_{ij}-\lambda \delta_{ij})c_{j}=0.$$
Here $\delta_{ij}$ satisfies $\delta_{ij}=0$ for $i\neq j$ and
$\delta_{ij}=1$ for $i=j$. Such a system has solutions which are
different from zero only if
\begin{equation}
|O_{ij}-\lambda \delta_{ij}|=0. \end{equation} Eq.(54) is called
fractional characteristic equation (or fractional secular equation)
which enables us to determine all the eigenvalues of the operator
$\hat{A_{\alpha}}$, that is, its spectrum.

\section{Conclusion }

In this paper, we propose anomalous quantum mechanics from operator
method.  We first introduce anomalous operators with fractional
derivative and exponential quantum associated with power observables
in coordinate space and in momentum
 space, respectively. Using the definition of anomalous operator and
 the exponential hypothesis that in heterogeneous complex media the
energy $E$ and momentum $p$ of a particle of mass $m$ satisfy
$E=\frac{p^{2\alpha}}{(2m)^{\alpha}}$,
 we then derived the fractional Schr$\ddot{\texttt{o}}$dinger equation (22) for the free particle
and (23) in the potential $V(x^{\beta})$. Besides, we derive the
fractional equation (32) of motion  for  anomalous quantum
mechanics. Furthermore, we obtain the fractional virial equations
(34) and (35) from which we obtain the anomalous hydrogen atom whose
transition energy values are much
 higher than that for the Bohr hydrogen atom and  match up the
anomalous $X-$ray spectrums observed by Wang, etc.. Finally, we give
the anomalous Heisenberg picture which equivalent to the fractional
Schr$\ddot{\texttt{o}}$dinger equation. There are so many studies
that one does not done the anomalous quantum mechanics in
relativistic case from operator method and others.

\section{Acknowledgements}

This work was supported by
the National Natural Science Foundation of China (No. 11626047) and
the Foundation for Young Key Teachers of Chengdu University of
Technology, China (Grant No. KYGG201414).

\end{document}